\documentclass[12pt]{llncs}
\usepackage{times}
\usepackage{xspace}
\usepackage{tikz}
\usepackage{graphicx}
\usepackage{color}
\usepackage[boxed]{algorithm2e}

\newcommand{\sq}{\hbox{\rlap{$\sqcap$}$\sqcup$}}
\renewcommand{\qed}{\hspace*{\fill}\sq}
\renewenvironment{proof}{\noindent {\bf Proof.}\ }{\qed\par\vskip 4mm\par}

\newcommand{\G}{\ensuremath{\cal{G}}\xspace}

\newcommand{\removed}[1]{}

\begin{document}

\title{Energy and Time Efficient Scheduling of Tasks with Dependencies 
on Asymmetric Multiprocessors 
\thanks{This work has been partially supported by the IST Programme of the European Union under contract number
IST-2005-15964 (\textsf{AEOLUS}) and by the ICT Programme of the European Union under contract number ICT-2008-215270 (\textsf{FRONTS}).} 
\thanks{An early version of some of the ideas of our work will appear as a \textit{brief announcement} in the 27$^{th}$ ACM Symposium on Principles of Distributed Computing (PODC 2008).}}

\author{Ioannis Chatzigiannakis\inst{1,2} \and Georgios Giannoulis\inst{2} \and Paul G. Spirakis\inst{1,2}}

\institute{Research Academic Computer Technology Institute (CTI), 
P.O.~Box 1382, N.~Kazantzaki Str., 26500 Patras, Greece 
\and Department of Computer Engineering and Informatics (CEID),  
University of Patras, 26500, Patras, Greece\\
Email: \email{\{ichatz,spirakis\}@cti.gr, giannulg@ceid.upatras.gr}}


\date{}

\maketitle


\begin{abstract}
In this work we study the problem of scheduling tasks with dependencies in multiprocessor architectures where processors have different speeds.
We present the preemptive algorithm ``Save-Energy" that given a schedule of tasks it post processes it to improve the energy efficiency without any deterioration of the makespan. In terms of time efficiency, we show that preemptive scheduling in an asymmetric system can achieve the same or better optimal makespan than in a symmetric system. Motivited by real multiprocessor systems, we investigate architectures that exhibit limited asymmetry: there are two essentially different speeds. Interestingly, this special case has not been studied in the field of parallel computing and scheduling theory; only the general case was studied where processors have $K$ essentially different speeds. We present the non-preemptive algorithm ``Remnants'' that achieves almost optimal makespan. We provide a refined analysis of a recent scheduling method. Based on this analysis, we specialize the scheduling policy and provide an algorithm of $\left(3 + o(1) \right)$ expected approximation factor. Note that this improves the previous best factor (6 for two speeds). We believe that our work will convince researchers to revisit this well studied scheduling problem for these simple, yet realistic, asymmetric multiprocessor architectures.
\end{abstract}

\section{Introduction}
\label{sec:Intro}

It is clear that processors technology is undergoing a vigorous shaking-up to allow one processor socket to provide access to multiple logical cores. Current technology already allows multiple processor cores to be contained inside a single processor module. Such chip multiprocessors seem to overcome the thermal and power problems that limit the performance that single-processor chips can deliver. Recently, researchers have proposed multiprocessor platforms where individual processors have different computation capabilities (e.g., see \cite{GDW06}). Such architectures are attractive because a few high-performance complex processors can provide good serial performance, and many low-performance simple processors can provide high parallel performance. Such asymmetric platforms can also achieve energy-efficiency since the lower the processing speed, the lower the power consumption is~\cite{WWDS96}. Reducing the energy consumption is an important issue not only for battery operated mobile computing devices but also in desktop computers and servers.

As the number of chip multiprocessors is growing tremendously, the need for algorithmic solutions that efficiently use such platforms is increasing as well. In these platforms a key assumption is that processors may have different speeds and capabilities but that the speeds and capabilities do not change. We consider multiprocessor architectures $P = \{ P_k: k = 1, \ldots, m\}$, where $c(k)$ is the speed of processor $p_k$. The total processing capability of the platform is denoted by $p = \sum_{k=1}^m c(k)$.

One of the key challenges of asymmetric computing is the scheduling problem. Given a parallel program of $n$ tasks represented as a dependence graph, the scheduling problem deals with mapping each task onto the available asymmetric resources in order to minimize the makespan, that is, the maximum completion time of the jobs. In this work we also look into how to reduce energy consumption without affecting the makespan of the schedule. Energy efficiency for speed scaling of parallel processors, which is not assumed in this work, was considered in \cite{AMS07}.

Our notion of a parallel program to be executed is a set of $n$ tasks  represented by $\G = (V,E)$, a Directed Acyclic Graph (DAG). The set $V$ represents $n=|V|$ simple tasks each of a \textit{unit} processing time. If task $i$ precedes task $j$ (denoted via $i \prec j$), then $j$ cannot start until $i$ is finished. The set $E$ of edges represents precedence constraints among the tasks. We assume that the whole DAG is presented as an input to the multiprocessor architecture. Our objective is to give schedules that complete the processing of the whole DAG in small time. Using terminology from scheduling theory, the problem is that of scheduling precedence-constrained tasks on related processors to minimize the makespan. In our model the \textit{speed asymmetry} is the basic characteristic. We assume that the overhead of (re)assigning processors to tasks of a parallel job to be executed is \textit{negligible}. 

The special case in which the DAG is just a collection of chains is of importance because general DAGs can be scheduled via a maximal chain decomposition technique of \cite{CB01}. Let $L = \{L_1, L_2, \ldots, L_r\}$ a program of $r$ Chains of tasks to be processed. We denote the length of chain $L_i$ by $l_i=|L_i|$, the count of the jobs in $L_i$; without loss of generality $l_1 \geq l_2 \geq \ldots \geq l_r$. Clearly $n=\sum^{r}_{i=1}l_i$.
In this case the problem is also known as \textit{Chains Scheduling}. Note that the decomposition technique of \cite{CB01} requires $O\left(n^3\right)$ time and the maximal chain decomposition depends only on the jobs of the given instance and is independent of the machine environment.

Because the problem is NP-hard \cite{U75} even when all processors have the same speed, the scheduling community has concentrated on developing approximation algorithms for the makespan. Early papers introduce $O(\sqrt{m})$-approximation algorithms \cite{HLS77,J80}, and more recent papers propose $O(\log{m})$-approximation algorithms \cite{CS99,CB01}. Numerous asymmetric processor organizations have been proposed for power and performance efficiency, and have investigated the behavior of multi-program\-med single threaded applications on them. \cite{BRUL05} investigate the impact of performance asymmetry in emerging multiprocessor architectures. They conduct an experimental methodology on a multiprocessor systems where individual processors have different performance. They report that the running times of commercial application may benefit from such performance asymmetry.

Previous research assumed the general case where multiprocessor platforms have $K$ distinct speeds. Yet recent technological advances (e.g., see \cite{GDW06,WWDS96}) build systems of two essential speeds.  Unfortunately, in the scheduling literature, the case of just $2$ distinct processor speeds has not been given much attention. In fact, the best till now results of \cite{CB01} reduce instances of arbitrary (but related) speeds, to at most $K = O(\log m)$ distinct speeds. Then the same work gives schedules of a makespan at most $O(K)$ times the optimal makespan, where $O(K)$ is $6K$ for general DAGs. We consider architectures of chip multiprocessors consisting of $m$ processors, with $m_s$ fast processors of speed $s>1$ and of $m-m_s$ slow processors of speed $1$ where the energy consumption per unit time is a convex function of the processor speed. Thus, our model is a special case of the uniformly related machines case, with only two \textit{distinct} speeds. 
In fact, the notion of distinct speeds used in \cite{CS99} and \cite{CB01} allows several speeds for our model, but not differing much from each other. So for the case of 2 speeds, considered here, this gives a $12$-factor approximation for general DAGs. Our goal here is to improve on this and under this simple model provide schedules with better makespan. We also focus on the special case where the multiprocessor system is composed of a single fast processor and multiple slow processors, like the one designed in \cite{GDW06}. Note that \cite{BP07} has recently worked on a different model that assumes asynchronous processors with time varying speeds.

\section{Energy Efficiency of Scheduling on Asymmetric Multiprocessors}
\label{sec:Energy}

Asymmetric platforms can achieve energy-efficiency since the lower the processing speed, the lower the power consumption is~\cite{WWDS96}. Reducing energy consumption is important for battery operated mobile computing devices but also for desktop computers and servers. To examine the energy usage of multiprocessor systems we adopt the model of \cite{AMS07}: the energy consumption per unit time is a convex function of the processor speed. In particular, the energy consumption of processor $k$ is proportional to $c(k)^{\alpha} \cdot t$, where $\alpha > 1$ is a constant. Clearly by increasing the makespan of a schedule we can reduce the energy usage.

We design the \textit{preemptive algorithm ``Save-Energy''} (see Alg.\ref{algo:saveenergy}) that post processes a schedule of tasks to processors in order to improve the energy efficiency by reassigning tasks to processors of slower speed. We assume no restrictions in the number of speeds of the processors and rearrange tasks so that the makespan is not affected. This reduces the energy consumption since in our model the energy spent to process a task is proportional to the speed of the processor to the power of $\alpha$ (where $\alpha >1$). In this sense, our algorithm will optimize a given schedule so that maximum energy efficiency is achieved.

\begin{algorithm}
{\SetLine \dontprintsemicolon 
\KwIn{An assignment of tasks to processors} 
\KwOut{An assignment of tasks to processors with reduced energy consumption} 
\BlankLine

Split schedule in intervals $t_j$, where $j\in[1\ldots \tau_0]$\;
Sort times in ascending order.\; $\tau \leftarrow \tau_0$\;
    \For{$c = c(2)$ \KwTo $c(m)$}{
        \For{$i \leftarrow 1$ \KwTo $\tau$}{
            $H \mbox{ holes in lower speeds that processing of } \ t_i \ \mbox{can fit without conflict in other assignments}$ \;
                Fit task $h$ in as many slower speeds starting from holes at c(1) to c, but if at $\tau_i$, $h$ fits to 2 or more speeds fill the hole closest to $\sqrt[\alpha-1]{\frac{1}{\alpha}}\cdot c$\;
                \If{$h$ does not fit exactly}{
                    Create a new $t'$ at the time preemption happens\;
                    Fit $h$ in extended slot\;
            }\;
        }\;
        $\tau \leftarrow \tau_{(previous)} + { \mbox{ Set of times that preemption occured} }$
    }\;
} \caption{{\label{algo:saveenergy} ``Save-Energy"}}
\end{algorithm}

We start by sorting the processors according to the processing capability $p_1, \ldots, p_m$ so that $c(1) \geq c(2) \ldots \geq c(m)$. We then split time in intervals $t_j$, where $j\in[1\ldots \tau_0]$, where $\tau_0$ is such that between these intervals there is not any preemption, no task completes and no changes are made to the precedence constraints. Furthermore we denote $x_{i}^j = 1$ if at $t_j$ we use $c(i)$ and $0$ otherwise. So the total energy consuption of the schedule is $E = \sum_{i=1}^{m}{\sum_{j=1}^{\tau_0}}x_{i}^j c(i)^{\alpha} t_j$.
%

\begin{theorem}[Condition of optimality]{\rm
If $E$ is the optimal energy consumption of a schedule (i.e., no further energy savings can be achieved), the following holds: There does not exist any $t_i, t_j$, where $i,j\in [1\ldots \tau_0]$, so that a list $l$ initially assigned to speed $c(u)$ at time $t_i$ can be rescheduled to $t_j$ with speed $c(v) \neq c(u)$ and reduce energy.}
\end{theorem}
\begin{proof}
Suppose that we can reduce the energy $E$ of the schedule. We obtain a contradiction. We can assume without any loss of generality that there exists at time $t_i$ a core $u$ that processes a list at speed $c(u)$ and there is a $t_j$ so that we can reschedule it to processor $v$ with speed $c(v) < c(u)$. This is so because if $c(v) \ge c(u)$ we will not have energy reduction. Therefore since $t_i, t_j$ exists then the new energy $E'$ must be lower than $E$. There exist only three cases when we try to reschedule a list $l$ from $t_i$ to $t_j$ from $c(u)$ to $c(v)$ where $c(v) < c(u)$:\\
\begin{itemize}

\item[(1)] \textit{The process of $l$ at $t_i$ fits exactly to $t_j$}. This is the case when $t_i \cdot c(u) = t_j \cdot c(v)$. In this case $E' = E - c(u)^{\alpha}t_i + c(v)^{\alpha}t_j$. But this violates the requirement $E' < E$ since $\frac{c(v)^{\alpha}t_j}{c(u)^{\alpha}t_i}<1$ because $\frac{c(v)^{\alpha}}{c(u)^{\alpha}}\cdot
\frac{t_j}{t_i}=\frac{c(u)^{\alpha}}{c(v)^{\alpha}} \cdot \frac{c(u)}{c(v)}=\left(\frac{c(u)}{c(v)}\right)^{\alpha-1}<1$ (recall that $\alpha>1$).   \\

\item[(2)] \textit{The process of $l$ at $t_i$ fits to $t_j$ and there remains time at $t_j$}. This is the case when $t_i \cdot c(u) < t_j \cdot c(v)$. Again we reach a contradiction since the new energy is the same with the previous case since $t_i \cdot c(u) < t_j \cdot c(v)$ and there exists $t_j'$ so that $t_i \cdot c(u)=t_j' \cdot c(v)$. \\

\item[(3)] \textit{The process of $l$ at $t_i$ does not fit completely to $t_j$}. This is the case when $t_i \cdot c(u) > t_j \cdot c(v)$. Now we cannot move all the processing of $l$ from $t_i$ to $t_j$. So there exists $t_i'$ so that $c(u) \cdot t_i' = c(v) \cdot t_j$. So the processing splits in two, at time $t_i$ for $t_i-t_i'$ and completely to $t_j$. The energy we save is
$c(u)^{\alpha} \cdot \left(t_i-t_i' \right) + c(v)^{\alpha} \cdot t_j - c(u)^{\alpha} \cdot t_i = c(v)^{\alpha} \cdot t_j - c(u)^{\alpha} \cdot t_i' <0$
because $\frac{c(v)}{c(u)}<1 \Rightarrow \frac{c(v)^{\alpha}}{c(u)^{\alpha}}\cdot\frac{c(u)}{c(v)}<1\Rightarrow
\frac{c(v)^{\alpha}}{c(u)^{\alpha}} \cdot \frac{t_j}{t_i} < 1$ which proves the theorem.
\end{itemize}
\end{proof}

\begin{theorem}{\rm
If the processing of list $l$ at $t_i$ at speed $c(u)$ fits completely to $t_j$ to two different speeds or more, we save more energy if we reschedule the list to the speed which is closer to 
$\sqrt[\alpha-1]{\frac{1}{\alpha}}\cdot c(u) $, and when $\alpha$ is $2$ it simplifies to $\frac{c(u)}{2}$.  }
\end{theorem}
\begin{proof}
The whole processing of $l$ must not change. So ${t'_i}$ is the time that the list will remain on speed $c(u)$ and can be calculated by the equation ${t'}_i \cdot c(u) + t_j \cdot c(v) = t_i \cdot c(u)$. So the energy that we spend if we do not use $c(v)$ is $E_{start}=c(u)^{\alpha} \cdot {t}_i$ and if we use $c(v)$ is
$E_{c(v)}=c(u)^{\alpha} \cdot {t'}_i + c(v)^{\alpha} \cdot t_j$.

So $E_{c(v)}={t'}_i \cdot c(u)^{\alpha}+ t_j \cdot c(v)^{\alpha} = E_{start} - c(v) \cdot t_j \left(c(u)^{\alpha-1}-c(v)^{\alpha-1} \right)$ and
because $c(u)>c(v)$ we always save energy if we reschedule any task
to a lower speed. The minimum energy occurs when the differential
equals to zero. That happens when $\left(t_j \cdot c(v) \cdot c(u)^{\alpha-1} -
t_j \cdot c(v)^{\alpha} \right)' = 0 \Rightarrow t_j \cdot c(u)^{\alpha-1} =\alpha
t_j \cdot c(v)^{\alpha-1} \Rightarrow c(v) =
\sqrt[\alpha-1]{\frac{1}{\alpha}}\cdot c(u)$. Now if the fragment of
the list can be reassigned to a further smaller $c(i)$ we obtain an
even smaller energy schedule. Thus we try to fill all the holes
starting from lower speeds and going upwards, in order to prevent
total fragmentation of the whole schedule and obtain a schedule of
nearly optimal energy consumption on the condition of unharmed
makespan.
\end{proof}

The algorithm ``Save-Energy" clearly does not increase the makespan since it does not delay the processing of any task, instead there may be even a reduction of the makespan. The new hole has size $\frac{c(v)}{c(u)}<1$ of the previous size and in every execution, a hole that can be filled goes to a faster processor. 

In arbitrary DAGs the problem is that due to precedence constraints we cannot swap two time intervals. To overcome this problem we proceed as follows: we define the supported set $(\textsf{ST}_i)$ to be all the tasks
that have been completed until time $t_i$ as well as those currently running and those who are ready to run. Between two intervals that have the same \textsf{ST} we can swap, or reschedule any assignment so we run
the above algorithm between all of these marked time intervals distinctly to create local optimums. In this case the complexity of the algorithm reduces to $O(m^2\cdot \sum\theta_i^2)$ where $\theta_i$ is the time between two time intervals with different \textsf{ST} while in list of tasks the time complexity is $O(\tau_0^2\cdot m^2)$. We note that in general DAGs the best scheduling algorithms for distinct speeds produces an $O(\log K)$-approximation (where $K$ is the number of essential speeds). In cases where schedules are far from
tight, the energy reduction that can be achieved in high.

\section{Time Efficiency of Scheduling on Asymmetric Multiprocessors}

We continue by providing some arguments for using asymmetric multiprocessors in terms of time efficiency. We show that preemptive scheduling in an asymmetric multiprocessor platform achieves the same or better optimal makespan than in a symmetric multiprocessor platform. The basic characteristic of our approach is \textit{speed asymmetry}. We assume that the overhead of (re)assigning processors to tasks of a parallel job to be executed is \textit{negligible}.
\begin{theorem}\label{LISToptimal}{\rm
Given any list $L$ of $r$ chains of tasks to be scheduled on preemptive machines, an asymmetric multiprocessor system will always have a better or equal optimal makespan than a symmetric one, given that both have the same average speed ($s'$) and the same total number of processors ($m$). The equality holds if during the whole schedule all processors are busy.}
\end{theorem}
\begin{proof}
Again we start by sorting the processors according to the processing capability $p_1, \ldots, p_m$ so that $c(1) \geq c(2) \ldots \geq c(m)$. We then split time in intervals $t_j$, where $j\in\{1 \ldots m]$, so that between these $m$ intervals there is not any preemption, no task completes and no changes are made to the precedence constraints. This is feasible since the optimal schedule is feasible and has finite preemptions.

Let $OPT_\sigma$ the optimal schedule for the symmetric multiprocessor system. Now consider the interval $(t_i,t_{i+1})$ where all processors process a list and divide it in $m$ time intervals. We assign each list to each of the $m$ asymmetric processors that are active, so that a task is assigned sequentially to all processors in the original schedule of $OPT_\sigma$. So each task will be processed by any processor for $\frac{1}{m} \cdot (t_{i+1}-t_i)$ time. Thus every task will have been processed during $(t_{i},t_{i+1})$ with an average speed of $\frac{\sum_{i=1}^m c(i)}{m}$, which is the speed of every symmetric processor. Thus given an optimal schedule for the symmetric system we can produce one that has at most the same makespan on the asymmetric set of processors.

The above is true when all processors are processing a list, at all times. Then the processing in both cases is the same. Of course there are instances of sets of lists that cannot be made to have all processors running at all times. In such schedules the optimal makespan on the asymmetric platform is better. Recall that we have sorted all speeds. Since the system is asymmetric it must have at least $2$ speeds. If at any time of $OPT_\sigma$ we process less lists than processors, following the analysis above, we will have to divide the time in $(\mbox{number of lists processing})<m$ (denoted by $\lambda$). So during time-interval $(t_i,t_{i+1})$ the processing of any list that is processed on symmetric systems will be $s' \cdot (t_{i+1}-t_i)$. While for the asymmetric system, the processing speed for the same time-interval will be $\frac{\sum_{i=1}^\lambda c(i)}{\lambda}$. Note that sum in the second equation is bigger than that of the first. That is valid because we use only the fastest processors. More formally $\frac{c(1)}{1}\geq\frac{c(1)+c(2)}{2}\geq\ldots\frac{c(1)+\ldots+c(\lambda)}{\lambda}\geq \ldots>\frac{c(1)+\ldots+c(m)}{m}$. So we produced a schedule that has a better makespan than $OPT_\sigma$. In other words, if during the optimal schedule for a symmetric system there exists at least one interval where a processor is idle, we can produce an optimal schedule for the asymmetric multiprocessors platform with smaller makespan.
\end{proof}

\begin{theorem}\label{DAGoptimal}{\rm
Given any DAG \G of tasks to be scheduled on preemptive machines, an asymmetric multiprocessor system will always have a better or equal optimal makespan than a symmetric one, provided that both have the same average speed ($s'$) and the same total number of processors ($m$). The equality holds if during the whole schedule all processors are busy.}
\end{theorem}
\begin{proof}
We proceed as above. The difference is that we split time in $(t_1,t_2,\ldots,t_m)$ that have the following
property: between any of these times $(t_i,t_{i+1})$ there is not any preemption on processors or completion of a \textit{list} or support for any \textit{list} that we could not process at $t_i$ due to precedence-constraints.

When all processors are processing a list, at all times, the processing in both asymmetric and symmetric systems is the same, i.e., $m \cdot s' \cdot (t_{i+1}-t{i})$. Of course there are DAGs that cannot be made to have all processors running at all times due to precedence-constraints or due to lack of tasks. In such DAGs the optimal makespan on the assymetric system is better than that of the symmetric one. If at any time of $OPT_\sigma$ we process less lists than processors, following the analysis of Theorem \ref{LISToptimal} we have that on the symmetric system the total processing will be $\sum_{j=1}^{\lambda} s' \cdot (t_{i+1}-t_i) = \lambda\cdot s'\cdot (t_{i+1}-t_i)$ while the processing speed during the same interval on the asymmetric one will be $\sum_{i=1}^\lambda c(i)\cdot (t_{j+1}-t_j)$ which is better.
\end{proof}

\section{Multiprocessor Systems of Limited Asymmetry}

We now focus on the case where the multiprocessor system is composed of a single fast processor and multiple slow ones, like the one designed in \cite{GDW06}. Consider that the fast processor has speed $s$ and the remaining $m-1$ processors have speed $1$. In the sequel preemption of tasks is not allowed.

We design the \textit{non-preemptive algorithm ``Remnants''} (see Alg.\ref{algo:remnants}) that always gives schedules with makespan $T \leq T_{opt} + \frac{1}{s}$. We greedily assign the fast processor first in each round. Then we try to maximize parallelism using the slow processors. In the beginning of round $k$ we denote $rem_k(i)$ the \textit{suffix} of list $L_k$ not yet done. Let $R_k(i)=|rem_k(i)|$. For $n$ tasks, the algorithm can be implemented to run in $O\left(\frac{1}{s}n^2\log{n} \right)$ time. The slow processors, whose ``list" is taken by the speedy processor in round $k$, can be reassigned to free remnants. Remark in the speed assignment produced by ``Remnants'' we can even \textit{name} the processors assigned to tasks (in contrast of general speed assignment methods, see e.g., \cite{J80,CS99,CB01}). Thus the actual scheduling of tasks is much more easy and of reduced overhead.

\begin{algorithm}
{\SetLine \dontprintsemicolon
\KwIn{Lists $L_1, \ldots, L_r$ of tasks}
\KwOut{An assignment of tasks to processors}
\BlankLine

$k\leftarrow 1$\;
\While{there are nonempty lists}{
    \lFor{$i \leftarrow 1$ \KwTo $r$}{$rem_k(i) = L_i$}\;
    $g_k \leftarrow \mbox{number of nonempty lists}$ \;
    Sort and rename the remnants so that $R_k(1) \geq R_k(2) \geq \ldots \geq R_k(g_k)$\;
    $u \leftarrow s, \ \ v \leftarrow 1$\;
    \tcc{Assign the fast processor sequentially to $s$ tasks}
    \While{$u > 0$ and $v \leq g_k$}{
        $p \leftarrow \min \left(u, R_k(v) \right)$\;
        Assign $p$ tasks of $rem_k(v)$ to fast processor and remove from $rem_k(v)$\;
        $u \leftarrow u - p, \ \ v \leftarrow v+1$\;
    }\;
    \tcc{Assign slow processors to beginning task of each remnant lists not touched by the fast speed assignment}
    \If{$v \leq g_k$}{
        $q \leftarrow \min(g_k, m-1)$\;
        \For{$w \leftarrow v$ \KwTo $q$}
        {
        Assign first task of $rem_k(w)$ to slow processor and remove from $rem_k(w)$\;
        }\;
    }\;
    Remove assigned tasks from the lists\;
    $k\leftarrow k + 1$\;
}\;
}
\caption{{\label{algo:remnants} ``Remnants''}}
\end{algorithm}

As an example, consider a system with 3 processors ($m=3$) where the speedy processor has $s=4$. In other words, we have a fast processor and two slow ones. We wish to schedule 4 lists, where $l_1=3, l_2=3, l_3=2$ and $l_4=2$. The ``remnants'' algorithm produces the following assignment with a makespan of $T=2$:
\begin{center}
\begin{tikzpicture}
\node[minimum size=0.35cm,circle,draw] (l11) at (0,0) {}
  [level distance=8mm]
  child[->] {node[minimum size=0.35cm,circle,draw] (l12) {}
   child[->] {node[minimum size=0.35cm,circle,draw] (l13) {}}
  };
\draw (l11) node[above=6pt] {{\small $L_1$}};

\node[minimum size=0.35cm,circle,draw] (l21) at (2,0) {}
  [level distance=8mm]
  child[->] {node[minimum size=0.35cm,circle,draw] (l22) {}
   child[->] {node[minimum size=0.35cm,circle,draw] (l23) {}}
  };
\draw (l21) node[above=6pt] {{\small $L_2$}};

\node[minimum size=0.35cm,circle,draw] (l31) at (4,0) {}
  [level distance=8mm]
  child[->] {node[minimum size=0.35cm,circle,draw] (l32) {}};
\draw (l31) node[above=6pt] {{\small $L_3$}};

\node[minimum size=0.35cm,circle,draw] (l41) at (6,0) {}
  [level distance=8mm]
  child[->] {node[minimum size=0.35cm,circle,draw] (l42) {}};
\draw (l41) node[above=6pt] {{\small $L_4$}};

\draw (l41) node[right=24pt,red] {Round 1};
\draw (l11) node[left=8pt] {{\tiny $s$}};
\draw (l12) node[left=8pt] {{\tiny $s$}};
\draw (l13) node[left=8pt] {{\tiny $s$}};
\draw (l21) node[left=8pt] {{\tiny $s$}};
\draw (l31) node[left=8pt] {{\tiny $1$}};
\draw (l41) node[left=8pt] {{\tiny $1$}};
\draw[red,line width=1.5pt] (-1,-2) .. controls (1,-2) .. (0.9,-0.8) .. controls (1,-0.3) .. (9,-0.3);

\draw (l42) node[right=24pt,red] {Round 2};
\draw (l22) node[left=8pt] {{\tiny $s$}};
\draw (l23) node[left=8pt] {{\tiny $s$}};
\draw (l32) node[left=8pt] {{\tiny $s$}};
\draw (l42) node[left=8pt] {{\tiny $s$}};
\draw[red,line width=1.5pt] (1,-2) .. controls (2.7,-2) .. (2.9,-1.5) .. controls (3.2,-1.2) .. (9,-1.2);

\end{tikzpicture}
\end{center}
Notice that the slow processors, whose ``list" is taken by the speedy processor in round $k$, can be reassigned to free remnants (one per free remnant). So our assignment tries to use all available parallelism per round. 

Now consider the case where the fast processor has $s=3$, that is, it runs slower than the processor of the above example. For the same lists of tasks, the algorithm now produces a schedule with a makespan of $T=2+\frac{1}{s}$:
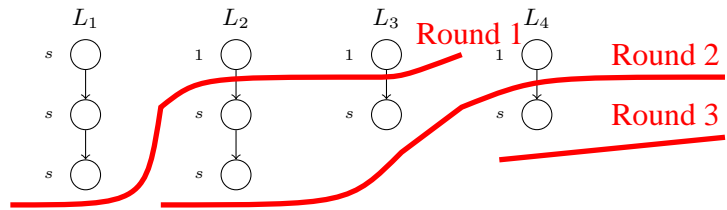
\begin{center}
\begin{tikzpicture}
\node[minimum size=0.35cm,circle,draw] (l11) at (0,0) {}
  [level distance=8mm]
  child[->] {node[minimum size=0.35cm,circle,draw] (l12) {}
   child[->] {node[minimum size=0.35cm,circle,draw] (l13) {}}
  };
\draw (l11) node[above=6pt] {{\small $L_1$}};

\node[minimum size=0.35cm,circle,draw] (l21) at (2,0) {}
  [level distance=8mm]
  child[->] {node[minimum size=0.35cm,circle,draw] (l22) {}
   child[->] {node[minimum size=0.35cm,circle,draw] (l23) {}}
  };
\draw (l21) node[above=6pt] {{\small $L_2$}};

\node[minimum size=0.35cm,circle,draw] (l31) at (4,0) {}
  [level distance=8mm]
  child[->] {node[minimum size=0.35cm,circle,draw] (l32) {}};
\draw (l31) node[above=6pt] {{\small $L_3$}};

\node[minimum size=0.35cm,circle,draw] (l41) at (6,0) {}
  [level distance=8mm]
  child[->] {node[minimum size=0.35cm,circle,draw] (l42) {}};
\draw (l41) node[above=6pt] {{\small $L_4$}};

\draw (l31) node[above=8pt,right=7pt,red] {Round 1};
\draw (l11) node[left=8pt] {{\tiny $s$}};
\draw (l12) node[left=8pt] {{\tiny $s$}};
\draw (l13) node[left=8pt] {{\tiny $s$}};
\draw (l21) node[left=8pt] {{\tiny $1$}};
\draw (l31) node[left=8pt] {{\tiny $1$}};
\draw[red,line width=2pt] (-1,-2) .. controls (0.8,-2) .. (1.0,-0.7)
                                  .. controls (1.5,-0.3) .. (3.8,-0.3)
                                  .. controls (4.2,-0.3) .. (5,-0);

\draw (l41) node[right=24pt,red] {Round 2};
\draw (l22) node[left=8pt] {{\tiny $s$}};
\draw (l23) node[left=8pt] {{\tiny $s$}};
\draw (l32) node[left=8pt] {{\tiny $s$}};
\draw (l41) node[left=8pt] {{\tiny $1$}};
\draw[red,line width=2pt] (1,-2) .. controls (3.5,-2) .. (4.2,-1.3)
                                 .. controls (4.2,-1.3) .. (5,-0.7)
                                 .. controls (6,-0.3) .. (8.5,-0.3);

\draw (l42) node[right=24pt,red] {Round 3};
\draw (l42) node[left=8pt] {{\tiny $s$}};
\draw[red,line width=2pt] (5.5,-1.4) .. controls (5.5,-1.4) .. (8.5,-1.1);

\end{tikzpicture}
\end{center}
Notice that for this configuration, the following schedule produces a makespan of $2$:
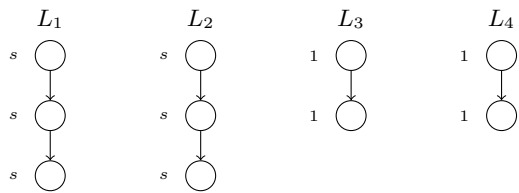
\begin{center}
\begin{tikzpicture}
\node[minimum size=0.35cm,circle,draw] (l11) at (0,0) {}
  [level distance=8mm]
  child[->] {node[minimum size=0.35cm,circle,draw] (l12) {}
   child[->] {node[minimum size=0.35cm,circle,draw] (l13) {}}
  };
\draw (l11) node[above=6pt] {{\small $L_1$}};

\node[minimum size=0.35cm,circle,draw] (l21) at (2,0) {}
  [level distance=8mm]
  child[->] {node[minimum size=0.35cm,circle,draw] (l22) {}
   child[->] {node[minimum size=0.35cm,circle,draw] (l23) {}}
  };
\draw (l21) node[above=6pt] {{\small $L_2$}};

\node[minimum size=0.35cm,circle,draw] (l31) at (4,0) {}
  [level distance=8mm]
  child[->] {node[minimum size=0.35cm,circle,draw] (l32) {}};
\draw (l31) node[above=6pt] {{\small $L_3$}};

\node[minimum size=0.35cm,circle,draw] (l41) at (6,0) {}
  [level distance=8mm]
  child[->] {node[minimum size=0.35cm,circle,draw] (l42) {}};
\draw (l41) node[above=6pt] {{\small $L_4$}};

\draw (l11) node[left=8pt] {{\tiny $s$}};
\draw (l12) node[left=8pt] {{\tiny $s$}};
\draw (l13) node[left=8pt] {{\tiny $s$}};
\draw (l21) node[left=8pt] {{\tiny $s$}};
\draw (l22) node[left=8pt] {{\tiny $s$}};
\draw (l23) node[left=8pt] {{\tiny $s$}};
\draw (l31) node[left=8pt] {{\tiny $1$}};
\draw (l32) node[left=8pt] {{\tiny $1$}};
\draw (l41) node[left=8pt] {{\tiny $1$}};
\draw (l42) node[left=8pt] {{\tiny $1$}};
\end{tikzpicture}
\end{center}
In the following theorem we show that the performance of Remnants is actually very close to optimal, in the sense of arguing that the above counter-example is essentially the only one.
\begin{theorem}
\label{Theorem4}{
For any set of lists $L$ and multiprocessor platform with one fast processor of speed $s$ and $m-1$ slow processors of speed $1$, if $T$ is the makespan of Algorithm Remnants then $T \leq T_{opt} + \frac{1}{s}$.}
\end{theorem}
\begin{proof}
We apply here the construction of Graham, as it was modified by \cite{CS99}, which we use in order to see if $T$
can be improved. Let $j_1$ a task that completes \textit{last} in Remnants. Without loss of generality, from the way Remnant works, we can always assume that $j_n$ was executed by the speedy processor. We consider now the logical chain ending with $j_1$ as follows: 
Iteratively define $j_{t+1}$ as a predecessor of $j_t$ that completes last of all predecessors of $j_t$
in Remnants. In this chain (a) either all its tasks were done at speed $c$ (in which case and since the fast processors works all the time, the makespan $T$ of Remnants is \textit{optimal}), or (b) there is a task $t^*$ at distance at most $s-1$ from $t_1$ that was done by speed $1$ in Remnants. In the later case, if $x$ is the start time of $t_1$, this means that before $x$ all speed $1$ processors are busy, else $t_1$ could be have scheduled earlier.
\begin{itemize}

\item[(b.1)]    If there is no other task in the chain done at speed $1$ and before $t_1$ then again $T$ is \textit{optimal} since before $t_1$ all processors of all speeds are busy.

\item[(b.2)] Let $t_2$ be another task in the chain done at speed $1$ and $t_2 < t_1$. Then $t_2$ must be an immediate predecessor of $t_1$ in a chain (because of the way Remnants work) and, during the execution of $t_2$, speed $s$ is busy but there could be some processor of speed $1$ available. Define $t_3, \ldots, t_j$ similarly (tasks of the last chain, all done in speed $1$ and $t_k < t_{k-1}$,  $k=j \ldots 2$). This can go up to the chain's start, which could have been done earlier by another speed $1$ processor and this \textit{is the only task} that could be done by an available processor, \textit{just one step before}. So, the makespan $T$ of algorithm Remnants can be compressed by only one task, and become optimal. But then $T\leq T_{opt}+\frac{1}{s}$ (i.e., it is the start of the last list that has no predecessor and which could go at speed $1$ together with nodes in the previous list).

\end{itemize}
\end{proof}

\subsection{An LP-relaxation approach for a schedule of good expected makespan}

In this section we relax the limitations to asymmetry. We work on the more general case of having $m_s$ fast processors of speed $s$ and $m-m_s$ slow processors of speed $1$. Note that we still have two distinct speeds and preemption of tasks is not allowed. We follow the basic ideas of \cite{CB01} and specialize the general lower bounds on makespan for the more general case. Clearly, the maximum \textit{rate} at which the multiprocessor system of limitted asymmetry can process tasks is $m_s\cdot s + (m-m_s)\cdot 1$,
which is achieved if and only if all machines are busy. Therefore to finish all $n$ tasks requires time at least
$A=\frac{n}{m_s \cdot s + 1 \cdot m-m_s}$. Now let
$$B=\max_{1 \leq j \leq \min(r,m)} \frac{\sum^j_{i=1}l_i}{\sum^j_{i=1}c(i)}$$
where
$c(1)=\ldots =c(m_s)=s$ and
$c(m_s+1)= \ldots = c(m) = 1$
are the individual processor speeds from the fast to the slow.
It follows that,
$$\frac{l_1}{s} \geq \frac{l_1+l_2}{2s} \geq \ldots \geq \frac{l_1+\ldots+l_j}{js} \qquad (j = m_s)$$
\textit{The interesting case is when $m_s<r$}.
So, we assume $m_s<r$ and let $l_s=l_1+l_2+\ldots +l_{m_s}$. Thus

$$B=\max_{m_s+1\leq j \leq \min(r,m)} \left(\frac{l_s+\sum^j_{i=m_s+1}l_i}{m_s\cdot (s-1)+j-1} \right)$$
By \cite{CB01} then
\begin{lemma}{\rm
Let $T_{opt}$ the optimal makespan of $r$ chains. Then
$T_{opt}\geq\max(A,B)$. }
\end{lemma}
Since the average load is also a lower bound for preemptive schedules we get
\begin{corollary}\label{corollary1}{\rm
$\max(A,B)$ is also a lower bound for preemptive schedules.}
\end{corollary}
As for the case where we have only one fast processor, i.e. $m_s=1$, in each step, at most $s + \min(m,r-1)$ tasks can be done since no two processors can work in parallel on the same list. This gives $T_{opt}\geq \frac{n}{s+\min(r-1,m)}$. Of course the bound $T_{opt}\geq B$ still also holds.

For a natural variant of list scheduling where no preemption takes place, called speed-based list scheduling, developed in \cite{CS99}, is constrained to schedule according to the speed assignments of the jobs. In classical list scheduling, whenever a machine is free the first available job from the list is scheduled on it. In this method, an available task is scheduled on a free machine
provided that the speed of the free machine matches the speed assignment of the task. The speed assignments of tasks have to be done in a clever way for good schedules. In the sequel, let $D_s = \frac{1}{s\cdot m_s}\cdot n_s$ where $n_s<n$ is the number of tasks assigned to speed $s$. Let $D_1=\frac{n-n_s}{m-m_s}$. Finally, for each chain $L_i$ and each task $j\in L_i$ with $c(j)$ being the speed assigned to $j$, compute $q_i=\sum_{j\in L_i}\frac{1}{c(j)}$ and let $C=\max_{i\in L}q_i $. The proof of the following theorem follows from an easy generalization of Graham's analysis of list scheduling.
\begin{theorem}[specialization of Theorem 2.1, \cite{CS99}]\label{Theorem1}{\rm
For any speed assignment ($c(j)=s$ or $1$) to tasks $j=1\ldots n$, the non-preemptive speed-based list scheduling method produces a schedule of makespan $T\leq C+D_s+D_1$.}
\end{theorem}

Based on the above specializations, we wish to provide a non-preemptive schedule (i.e., speed assignment) that achieves good makespan. We either assign tasks to speed $s$ or to speed $1$ so that $C+D_s+D_1$ is not too large. Let, for task $j$:

\begin{displaymath}
x_j=\left\{
\begin{array}{ll}
1 &\mbox{when}\ c(j)=s\\
0 &\mbox{otherwise}
\end{array}\right.
\end{displaymath}
\noindent and
\begin{displaymath}
y_j=\left\{
\begin{array}{ll}
1 &\mbox{when}\ c(j)=1\\
0 &\mbox{otherwise}
\end{array}\right.
\end{displaymath}
\noindent Since each task $j$ must be assigned to some speed we get
\begin{equation}\label{equ:1}
\forall j=1\ldots n \qquad x_j+y_j=1
\end{equation}
\noindent In time $D$, the fast processors can complete $\sum^n_{j=1}x_j$ tasks and the slow processors can complete $\sum^n_{j=1}y_j$ tasks. So
\begin{equation}\label{equ:2}
\frac{\sum^n_{j=1}x_j}{m_s\cdot s}\leq D
\end{equation}
\noindent and
\begin{equation}\label{equ:3}
\frac{\sum^n_{j=1}y_j}{m-m_s}\leq D
\end{equation}
\noindent Let $t_j$ be the completion time of task $j$
\begin{equation}\label{equ:4}
(t_j\geq 0)
\end{equation}
\noindent If $j'<j$ then clearly
\begin{equation}\label{equ:5}
\frac{x_j}{s}+y_j \leq t_j - t_{j'}
\end{equation}
\noindent Also
\begin{equation}\label{equ:6}
\forall j : t_j\leq D
\end{equation}
\noindent and
\begin{equation}
\label{equ:7}
\forall j : x_j,y_j \in \{ 0,1 \}
\end{equation}
Based on the above constraints, consider the following mixed integer program:

\begin{algorithm}[H]
\SetLine
\dontprintsemicolon
\textbf{MIP:}\;
$\min{D}$\;
under (\ref{equ:1}) to (\ref{equ:7})\;
\end{algorithm}

MIP's optimal solution is clearly a lower bound on $T_{opt}$.
Note that $(\ref{equ:2}) \Rightarrow D_s \leq D $
and $(\ref{equ:3}) \Rightarrow D_1 \leq D$.
Also note that since
$ t_{j'}\geq 0 \Rightarrow \frac{x_j}{s}+y_j\leq t_j$
by (\ref{equ:5}) and thus also $C\leq D$, by adding times on each chain. So, if we could solve
MIP then we would get a schedule of makespan $T\leq 3\cdot T_{opt}$,
by Theorem \ref{Theorem1}.
Suppose we relax (\ref{equ:7}) as follows:
\begin{equation}\label{equ:7-}
x_j,y_j\in[0,1]\qquad j=1\ldots n
\end{equation}
Consider the following linear program:

\begin{algorithm}[H]
\SetLine
\dontprintsemicolon
\textbf{LP:}\;
$\min{D}$\;
under (\ref{equ:1}) to (\ref{equ:7}) and (\ref{equ:7-})\;
\end{algorithm}

This LP can be solved in polynomial time and its optimal solution
$\overline{x_j},\overline{y_j},\overline{t_j}$, where  $j=1\ldots n$,
gives an optimal $\overline{D}$, also $\overline{D}\leq
T_{opt}$ (because $\overline{D}\leq \mbox{best $D$ of MIP}$).

We now use randomized rounding, to get a speed assignment $A_1$
$$
\forall task_j :
\begin{array}
{lcc} c(j') = s & \mbox{with probability} & \overline{x_j} \\
c(j)=1 & \mbox{with probability} & 1-\overline{x_j}=\overline{y_j}
\end{array}
$$
Let $T_{A_1}$ be the makespan of $A_1$. Since $T_{A_1} \leq C+D_s
+D_1 \Rightarrow E(T_{A_1})\leq E(C) + E(D_s) + E(D_1)$.
\noindent But note that
$$E(n_s) = \sum^n_{j=1}\overline{x_j} 
\qquad \qquad \mbox{and} \qquad \qquad 
E(n-n_s) = \sum^n_{j=1}\overline{y_j}$$
\noindent so $E(D_1), E(D_2) \leq D$ by (\ref{equ:2},\ref{equ:3}) and, for each list $L_i$
$$E\left( \sum_{j\in L_i}\frac{1}{c(j)} \right) =
\sum_{j\in L_i}E\left(\frac{1}{c(j)}\right) 
= \sum_{j\in L_i}\left(\frac{1}{s}\cdot \overline{x_j}+1\cdot \overline{y_j}\right)
\leq \overline{D} \qquad \mbox{by (\ref{equ:5}), (\ref{equ:6})}
$$
\noindent I.e., $E(C)\leq \overline{D}$. So we get the following theorem:
\begin{theorem}\label{Theorem2}{\rm
Our speed assignment $A_1$ gives a non-preemptive schedule of expected makespan at most $3\cdot T_{opt}$}
\end{theorem}
Our MIP formulation also holds for general DAGs and 2 speeds, when all tasks are of unit length. Since Theorem \ref{Theorem1} of \cite{CS99} and the lower bound of \cite{CB01} also holds for general DAGs, we get:
\begin{corollary}\label{Corollary 2}{\rm
Our speed assignment $A_1$, for general DAGs of unit tasks gives a non-preemptive schedule of expected makespan at most $3\cdot T_{opt}$.}
\end{corollary}
We continue by making some special consideration for lists of tasks, that is we think about DAGs that are decomposed in sets of lists. Then, $A_1$ can be greedily improved since all tasks are of unit processing time, as follows. After doing the assignment experiment for the nodes of a list $L_i$ and get $l_i^1$ nodes on the fast processors and $l_i^2$ nodes on slow processors. We then reassign the first $l_i^1$ nodes of $L_i$ to the fast processors and the remaining nodes of $L_i$ to the slow processors. Clearly this does not change any of the expectations of $D_s$, $D_1$ and $C$. Let $\widetilde{A_1}$ be this modified (improved) schedule. 

Also, because all tasks are equilenght (unit processing time), any reordering of them in the same list will not change the optimal solution of LP. But then, for each list $L_i$ and for each task $j\in L_i$, $\overline{x_j}$ is the same (call it $\overline{x_i}$), and the same holds for $\overline{y_j}$. Then the processing time of $L_i$ is just $\frac{f_i}{s} + \left(1 - f_i \right)$ where $f_i$ is as the Bernoulli $B(l_i,\overline{x_i})$.

In the sequel, let $\forall i : l_i \geq \gamma \cdot n$, for some $\gamma\in(0,1)$ and let $s \cdot m= o(n) = n^\epsilon$, where $\epsilon<1$. Then from $\widetilde{A_1}$ we produce the speed assignment $\widetilde{A_2}$ as follows:

\begin{algorithm}[H]
\SetLine
\dontprintsemicolon
\ForEach{list $L_i$, $i = 1 \ldots r$}{
  \eIf{$\overline{x_i}<\frac{\log{n}}{n}$}{
    \textit{assign all the nodes of $L_i$ to unit speed}\;
  }{
    \textit{for $L_i$, $\widetilde{A_2} = \widetilde{A_1}$}\;
  }
}
\end{algorithm}

Since for the makespan $T_{\widetilde{A_1}}$ of $\widetilde{A_1}$ we have
$$E \left( T_{\widetilde{A_1}} \right) = E \left( T_{A_1} \right) \leq 3T_{opt}$$
\noindent we get
$$E \left( T_{\widetilde{A_2}} \right) \leq 3\cdot T_{opt}+s\log{n}$$
\noindent But
$$T_{opt} \geq \frac{n}{s\cdot m_s + ( m-m_s )} \geq \frac{n}{sm}=n^{1-\epsilon}$$
\noindent Thus
$$E \left(T_{\widetilde{A_2}} \right) \leq \left(3+ o(1) \right)T_{opt}$$
\noindent However, in $\widetilde{A_2}$, the probability that
$T_{\widetilde{A_2}} > E \left(T_{\widetilde{A_2}} \right) \left(1 + \beta \right)$,
where $\beta$ is a constant $(0,1)$, is at most
$\frac{1}{\gamma}\exp{\left( -\frac{\beta^2}{2}\cdot l_i \cdot \overline{x_i} \right)}$
(by Chernoff bounds), i.e., at most
$\frac{1}{\gamma}{\left(\frac{1}{n}\right)}^{\frac{\beta^2}{2}}$.
This implies that it is enough to repeat the randomized assignment of speeds at most a polynomial number of times and get a schedule of actual makespan at most $\left(3+ o(1) \right)T_{opt}$. So, we get our next theorem:
\begin{theorem}
\label{Theorem3} {\rm
When each list has length $l_i\geq \gamma \cdot n$ (where $\gamma \in (0,1)$) and $s\cdot m = n^\epsilon$ (where $\epsilon<1$) then we get a (deterministic) schedule of actual makespan at most $\left(3+o(1) \right)T_{opt}$ in expected polynomial time. }
\end{theorem}

\section{Conclusions and Future work}
\label{sec:OurResults}

Processors technology is undergoing a vigorous shaking-up to enable low-cost multiprocessor platforms where individual processors have different computation capabilities. We examined the energy consumption of such asymmetric arcitectures. We presented the preemptive algorithm ``Save-Energy" that post processes a schedule of tasks to reduce the energy usage without any deterioration of the makespan. Then we examined the time efficiency of such asymmetric architectures.  We shown that preemptive scheduling in an asymmetric multiprocessor platform can achieve the same or better optimal makespan than in a symmetric multiprocessor platform.

Motivited by real multiprocessor systems developed in \cite{GDW06,WWDS96}, we investigated the special case where the system is composed of a single fast processor and multiple slow processors. We say that these architectures have limited asymmetry. Interestingly, alghough the problem of scheduling has been studied extensively in the field of parallel computing and scheduling theory, it was considered for the general case where multiprocessor platforms have $K$ distinct speeds. Our work attempts to bridge between the assumptions in these fields and recent advances in multiprocessor systems technology. In our simple, yet realistic, model where $K=2$, we presented the non-preemptive algorithm ``Remnants'' that achieves almost optimal makespan. 

We then generalized the limited asymmetry to systems that have more than one fast processors while $K=2$. We refined the scheduling policy of \cite{CS99} and give a non-preemptive speed based list Randomized scheduling of DAGs that has a makespan $T$ whose expectation $E(T)\leq 3\cdot OPT$. This improves the previous best factor (6 for two speeds).  We then shown how to convert the schedule into a deterministic one (in polynomial \textit{expected} time) in the case of long lists. 

Regarding future work we wish to examine trade-offs between makespan and energy and we also wish to investigate extensions for our model allowing other aspects of heterogeneity as well.


\end{document}